\documentclass[secnumarabic,amssymb,nobibnotes,aps,prl,preprint,letterpaper,preprintnumbers,amsmath,superscriptaddress,floatfix,notitlepage]{revtex4-1}
\usepackage[normalem]{ulem}
\usepackage{wrapfig}
\usepackage{amsmath}
\usepackage{enumitem}
\usepackage{multirow}
\usepackage[letterpaper,margin={1in,1in}]{geometry}
\usepackage{graphicx}
\usepackage{epstopdf}
\usepackage{lineno}
\usepackage{appendix}
\usepackage{longtable}
\usepackage{array}
\usepackage{siunitx}
\usepackage[dvipsnames]{xcolor}
\usepackage{lineno}
\setcitestyle{super}
\usepackage{multirow}
\DeclareSIUnit\mrad{\milli\rad}

\usepackage{soul}
\newcolumntype{L}{@{}>{\kern\tabcolsep}l<{\kern\tabcolsep}}
\usepackage{colortbl}
\usepackage{notes2bib}
\usepackage{marvosym}
\usepackage[colorinlistoftodos,prependcaption,textsize=tiny]{todonotes}

\begin{document}

\setlength{\LTcapwidth}{6.5in}

\title{Elementary excitations of single-photon emitters in hexagonal Boron Nitride}

\author{Jonathan Pelliciari \Letter}
\email{pelliciari@bnl.gov}
\affiliation{National Synchrotron Light Source II, Brookhaven National Laboratory, Upton, New York 11973, USA}

\author{Enrique Mejia}
\affiliation{Photonics Initiative, Advanced Science Research Center, City University of New York, New York City, NY 10031, USA}

\author{John M. Woods}
\affiliation{Photonics Initiative, Advanced Science Research Center, City University of New York, New York City, NY 10031, USA}

\author{Yanhong Gu}
\affiliation{National Synchrotron Light Source II, Brookhaven National Laboratory, Upton, New York 11973, USA}

\author{Jiemin Li}
\affiliation{National Synchrotron Light Source II, Brookhaven National Laboratory, Upton, New York 11973, USA}

\author{Saroj B. Chand}
\affiliation{Photonics Initiative, Advanced Science Research Center, City University of New York, New York City, NY 10031, USA}

\author{Shiyu Fan}
\affiliation{National Synchrotron Light Source II, Brookhaven National Laboratory, Upton, New York 11973, USA}

\author{Kenji Watanabe}
\affiliation{Research Center for Functional Materials, National Institute for Materials Science, Tsukuba, Ibaraki 305‑0044,
Japan. }

\author{Takashi Taniguchi}
\affiliation{International Center for Materials Nanoarchitectonics, National Institute for Materials Science, Tsukuba, Ibaraki 305‑0044, Japan. }

\author{Valentina Bisogni}
\affiliation{National Synchrotron Light Source II, Brookhaven National Laboratory, Upton, New York 11973, USA}

\author{Gabriele Grosso \Letter}
\email{ggrosso@gc.cuny.edu}
\affiliation{Photonics Initiative, Advanced Science Research Center, City University of New York, New York City, NY 10031, USA}
\affiliation{Physics Program, Graduate Center, City University of New York, New York, NY 10016, USA}

\date{\today}
\maketitle

{\bf  
Single-photon emitters serve as building blocks for many emerging concepts in quantum photonics. The recent identification of bright, tunable, and stable emitters in hexagonal boron nitride (hBN) has opened the door to quantum platforms operating across the infrared to ultraviolet spectrum. While it is widely acknowledged that defects are responsible for single-photon emitters in hBN, crucial details regarding their origin, electronic levels, and orbital involvement remain unknown.
Here, we employ a combination of resonant inelastic X-ray scattering and photoluminescence spectroscopy in defective hBN unveiling an elementary excitation at 285 meV that gives rise to a plethora of harmonics correlated with single-photon emitters. We discuss the importance of N $\pi^*$ antibonding orbitals in shaping the electronic states of the emitters. The discovery of the elementary excitations of hBN provides new fundamental insights into quantum emission in low-dimensional materials, paving the way for future investigations in other platforms.
}

\newpage

\begin{figure*}[h]
\centering
\includegraphics[width=\textwidth]{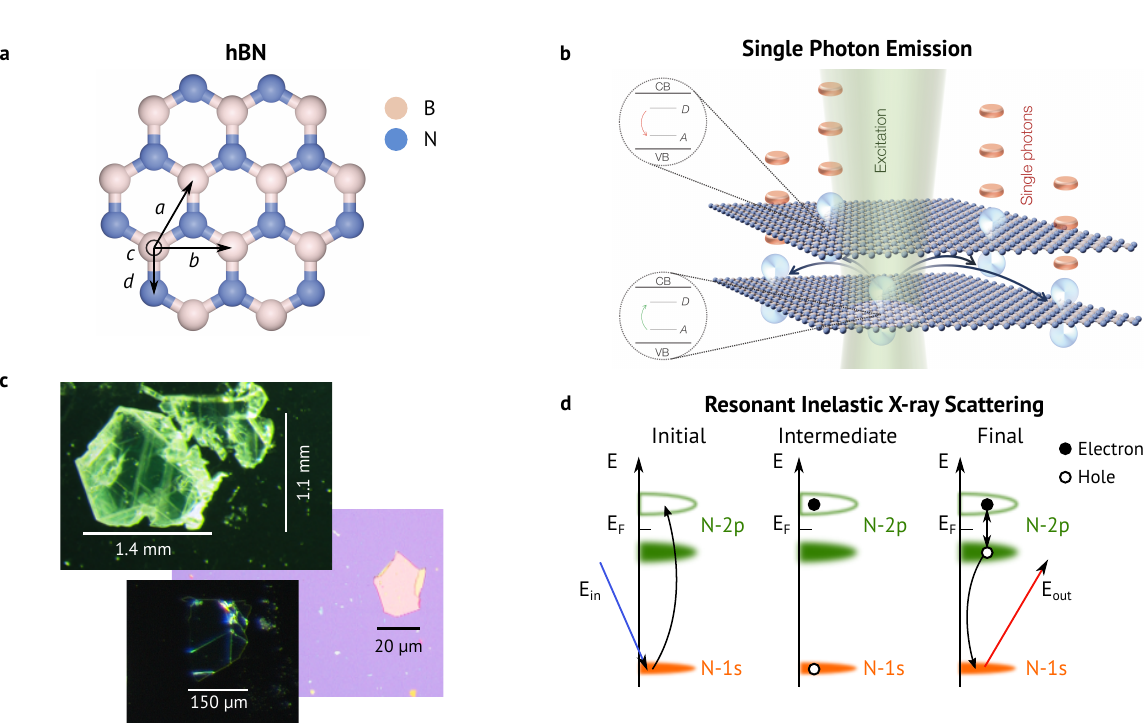}
\caption{ {\bf Single-photon emission and Resonant Inelastic X-Ray Scattering (RIXS) process in hBN.} {\bf a,} Crystal structure of a single layer of hBN. The lattice vector $\vec{c}$ points out of the plane and connect two hBN layers {\bf b,} Single photon emission process in layered hBN showing the formation of donor and acceptor energy states within the band-gap generated by defects. Upon excitation of a donor defect, electrons can recombine with acceptors at a distance of a few atomic sites and emit single photons. {\bf c,} Microscope images of hBN single crystals used for this study in bulk (top panel) and in thin flakes exfoliated on a Si/SiO$_2$ substrate (bottom panel). {\bf d,} Diagram of the generalized RIXS process at the N-K edge.}
\label{fig1}
\end{figure*}

Hexagonal boron nitride (hBN) is a versatile material with applications in multiple scientific fields due to its properties, including air stability, hyperbolic dispersion, and strong optical non-linearities \cite{caldwell_photonics_2019}. It is composed of an hexagonal lattice of alternating B and N atoms (Fig. \ref{fig1}\textbf{a}) whose hybridization leads to a wide band-gap in the ultraviolet range. The discovery of single photon emitters (SPEs) in hBN has spurred scientific interest due to their room temperature operation,\cite{tran_quantum_2016} high brightness \cite{bourrellier_bright_2016}, frequency tunability \cite{grosso_tunable_2017}, stability \cite{fournier_position-controlled_2021}, narrow linewidth \cite{dietrich2018observation}, and optically detected magnetic resonances \cite{chejanovsky2021single}.
In hBN, SPEs have been associated with defects whose electronic levels within the band gap generate quantum light with narrow and well defined energies (Fig. \ref{fig1}\textbf{b}) defined by the atomic composition of the orbitals involved in the electronic transition. Although these aspects have been extensively studied, no consensus to an underlying mechanism for hBN emitters has been reached, yet \cite{jungwirth_optical_2017,abdi_color_2018,tawfik_first-principles_2017,mendelson_identifying_2021}. The characterization of SPEs in hBN has been mostly limited to photoluminescence (PL) spectroscopy, measurements of the second order auto-correlation function ($g^2 (\tau)$), and magnetic resonances. These efforts identified the negatively charged boron vacancy ($V_{B}^{-}$) as a consistent defect\cite{gottscholl_2021_room,gottscholl_2020_initialization}, but could not reveal the properties of all other hBN emitters \cite{tran_2016_robust}, leaving the microscopic details necessary to explain the large distribution of the emission energies of SPEs in hBN still obscure. 

By combining resonant inelastic X-ray scattering (RIXS) and PL spectroscopy, we unveil the presence of elementary electronic excitations correlated to most of SPEs in the visible spectrum of hBN. Our RIXS measurements on defective hBN uncover an excitation at $E_{0}$ = 285 meV that propagates up to 2.3 eV through multiple regular harmonics $E_{n}=nE_{0}$ with $n \in \mathbb{N}$. The independent analysis of the PL spectral emission containing multiple SPE peaks with a model that includes recombination processes similar to donor-acceptor pairs (DAP) reveals fundamental transitions matching the energy of the harmonics observed with RIXS at $n= 5-8$. Thus, using only a limited set of parameters, we show that the sequence of harmonics $E_{n}$ based on the elementary energy $E_{0}$ can explain the wide frequency range of quantum emission in hBN.

\begin{figure*}[tbh]
\centering
\includegraphics[width=\textwidth]{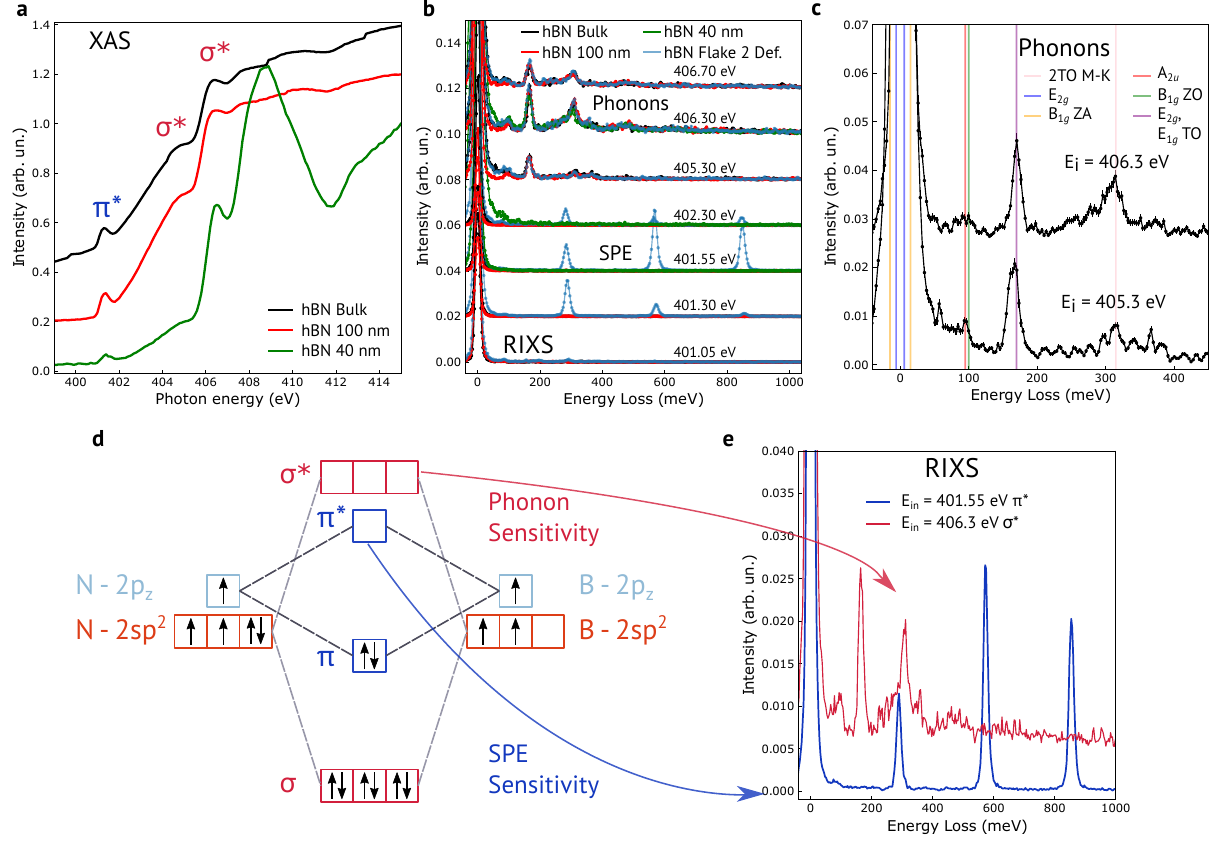}
\caption{ {\bf X-ray absorption spectroscopy (XAS) and Resonant Inelastic X-Ray Scattering (RIXS) on hBN}. {\bf a,} XAS of pristine hBN as a function of thickness in $\sigma$ polarization. {\bf b,} RIXS of pristine hBN and defective hBN as a function of thickness and incident energy. Incident polarization was set to $\sigma$. We can identify that at high incident energy (main N-K edge) the RIXS signal is qualitatively different than at the pre-edge. {\bf c,} Zoom in of the RIXS spectrum of an hBN crystal showing the sensitivity of RIXS to phonons. The energy of the main phonon modes extracted from Inelastic X-Ray Scattering and Raman studies \cite{serrano_vibrational_2007} are highlighted with vertical lines. {\bf d, e} Molecular orbital diagram of hBN ({\bf d}) that illustrates the connection between incident excitation energy and sensitivity of RIXS to phonons or single photon emitters ({\bf e}). Diagram assumes the same energy for N/B-2sp$^2$ and N/B 2p$_{z}$. The error bars of the RIXS intensity of panels \textbf{b,c,e} are based on a Poisson noise on the total intensity detected on the RIXS detector. When error bars are not visible they are smaller than the marker size.}
\label{fig2}
\end{figure*}

The sensitivity of RIXS to elementary electronic excitations is given by a resonance effect in which a core electron is excited to the conduction band creating a real short-lived (few femtoseconds) intermediate state that radiatively decays emitting a photon and leaving the system in a neutral excited state (Fig.~\ref{fig1}\textbf{d}) \cite{ament_resonant_2011}. The coherence of this process allows for sensitivity to all of the electronic degrees of freedom, \textit{i.e.} spin, charge, orbital, and lattice excitations \cite{ament_determining_2011,gelmukhanov_dynamics_2021}. Recent technological developments have enabled the study of micron-sized samples (Fig.~\ref{fig1}\textbf{c}) with high energy resolution \cite{dvorak_towards_2016,pelliciari_tuning_2021,pelliciari_evolution_2021, dean_spin_2012}. At the N-K edge, the atomic transition of RIXS is 1\textit{s}\textrightarrow2\textit{p/sp$^2$} and 2\textit{p/sp$^{2*}$}\textrightarrow1\textit{s} of nitrogen (as schematically highlighted in Fig.~\ref{fig1}\textbf{d}). 
In Fig.~\ref{fig2}\textbf{a} we report X-ray absorption spectra (XAS) at the N-K edge in $\sigma$ polarization for pristine hBN of different thickness (bulk, 100 nm, and 40 nm thick). The first peak at $\approx$401.5 eV is a transition from N$-1s$ to N$- \pi^*$ orbitals whereas the peaks at higher energy are ascribed to transitions from N$-1s$ to N$-\sigma^*$ orbitals \cite{mcdougall_influence_2017,vinson_resonant_2017,mcdougall_near_2014}
Pristine hBN samples are exfoliated flakes from high-quality bulk material with extremely low concentration of defects and no presence of SPEs or other emissions below the band gap.\cite{watanabe2004direct} 
Defective hBN samples are prepared by irradiating pristine flakes with argon plasma followed by high temperature annealing (see Methods). This technique has been proven to be effective in generating defect-based SPEs in hBN with sharp emission energies spanning a large spectral range \cite{xu_single_2018}. RIXS spectra at different incident photon energies for pristine as well as a defective hBN (20 nm) are reported in Fig. \ref{fig2}\textbf{b}. The RIXS spectra display a strong diffuse scattering line at 0 eV in energy loss and multiple peaks at finite energy. Depending on the incident energy, we identify two different responses. At the N$-\sigma^*$ XAS resonance, RIXS probes the phonon modes for both pristine and defective hBN, as revealed in Fig. \ref{fig2}\textbf{c}, Fig.~S7, and Fig.~S8. The measured energies for the phonon modes agree with Raman and inelastic X-ray scattering results for the bulk \cite{serrano_vibrational_2007}. At the N $-\pi^*$ XAS peak, no signal is observed for pristine hBN (Fig. \ref{fig2}\textbf{b}), whereas for defective hBN multiple harmonic peaks appear with an elementary energy of $E_{0}$ = 285 meV. The detection of phonons or harmonics as a function of incident photon energy is rationalized in Fig. \ref{fig2}\textbf{d,e} where we illustrate a diagram of the molecular orbitals of hBN (panel \textbf{d}). 
In hBN, the orbital hybridization leads to the formation of three $\sigma^*$ and one $\pi^*$ anti-bonding orbitals which, when excited in the RIXS process, result in phononic detection or the harmonics, respectively (Fig. \ref{fig2}\textbf{e}). In Fig. \ref{fig2}\textbf{e} we show that the energy of the harmonics does not match the one of the phonons, corroborating the different electronic sensitivity of RIXS rather than a mere rescaling of intensity of the same peaks. 

\begin{figure*}[tbh]
\centering
\includegraphics[width=\textwidth]{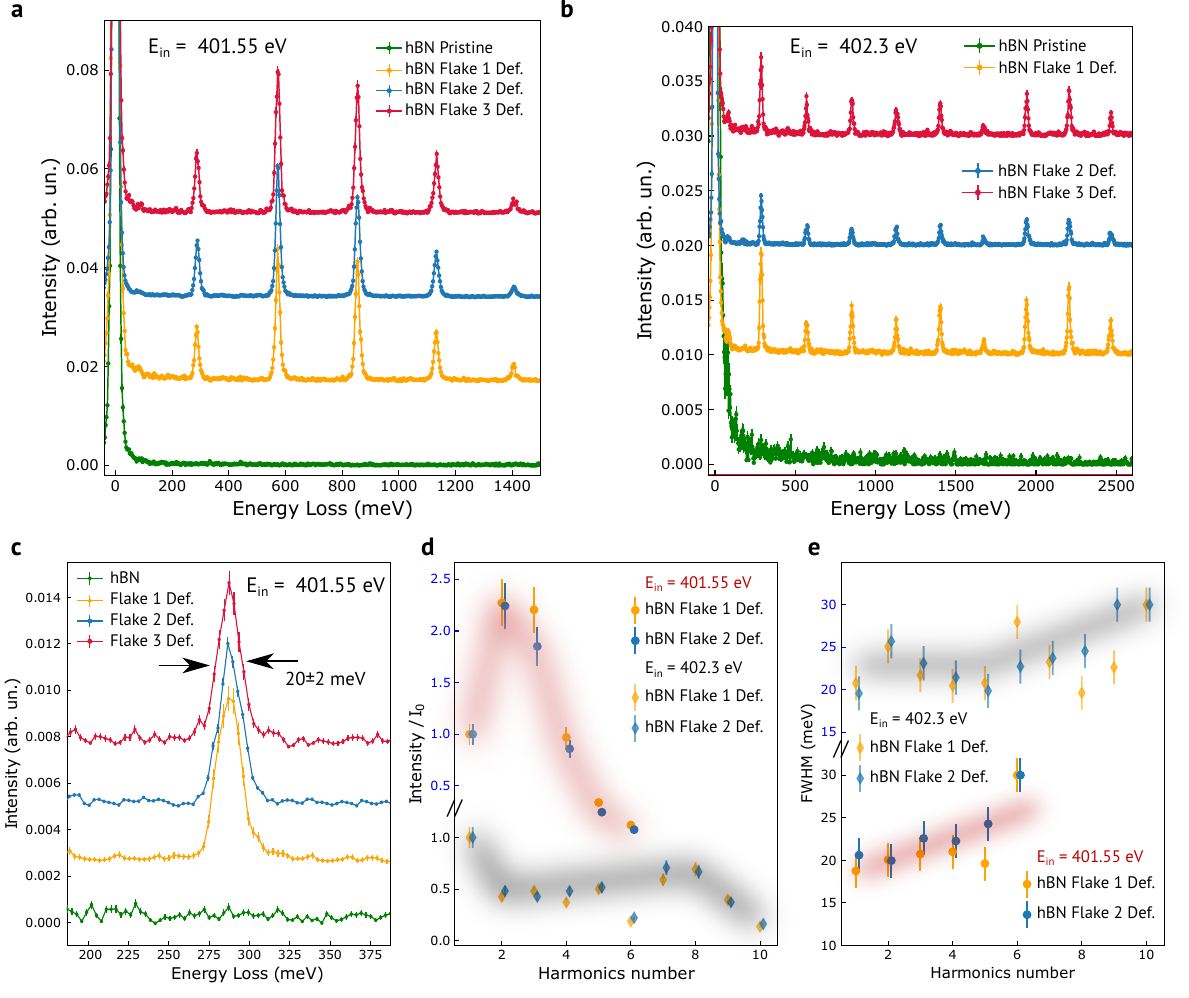}
\caption{ {\bf Fundamental excitations in defective flakes of hBN}. {\bf a,b} Resonant Inelastic X-Ray Scattering on thin  pristine hBN and defective flakes at E$_{in}$ of 401.55 and 402.3 eV. {\bf c,} Zoom in of the fundamental harmonics of defective hBN. The linewidth of the peak is comparable to the experimental resolution indicating that the peaks are extremely narrow in energy. 
{\bf d, e} Results of the Gaussian fit of the peaks as a function of harmonic index on defective flakes 1 and 2. In {\bf d} we report the intensity profile normalized by the intensity of the fundamental peak indicating a non-trivial dependence and a strong oscillator strength. In {\bf e} we summarize the FWHM of the peaks as a function of harmonic number. The FWHM changes very little over a wide range of energy loss spanning from the IR up to UV. The error bars of the RIXS intensity of panels \textbf{a-c} are based on a Poisson noise on the total intensity detected on the RIXS detector. When error bars are not visible they are smaller than the markers size. The error bars of panel \textbf{d} are based on an evaluation of the error associated with the fitting and errors due to the normalization. The error bars in panel \textbf{e} are based on the error on the fitting. }
\label{fig3}
\end{figure*}

We now discuss RIXS data from three defective hBN flakes of approximate thickness 20 nm (Flake 1), 100 nm (Flake 2) and 200 nm (Flake 3). Spectra in Fig.\ref{fig3}\textbf{a,b} are obtained by exciting at two energies around the N pre-edge $\pi^*$ transition. The series of harmonic peaks spans a different energy window depending on the incident photon energy: it extends up to 1.5 eV and 2.3 eV when exciting at 401.55 eV and 402.3 eV, respectively. The number of harmonics likely depends on the incident energy available to promote electrons above the Fermi level as the incident energy difference (0.750 eV) is consistent with the energy of the highest harmonic detected (1.5 vs 2.3 eV). The reproducibility on multiple samples is remarkable as plasma treatment is an aggressive and nonselective method to induce defects, indicating the robustness and generality of these harmonic in defective hBN. In Fig. \ref{fig3}\textbf{c}, we compare the first peak of the harmonic for three defective flakes showing that in all cases the peak linewidth approaches the RIXS energy resolution of 20 meV (instrumental broadening, see Fig.\ref{fig3}\textbf{e}). This further confirms that RIXS detects fundamental transitions in defective hBN, that their elementary energy is 285 meV, and that the harmonics are invariant under plasma treatment and flake thickness. 

\begin{figure*}[tbh]
\centering
\includegraphics[width=0.75\textwidth]{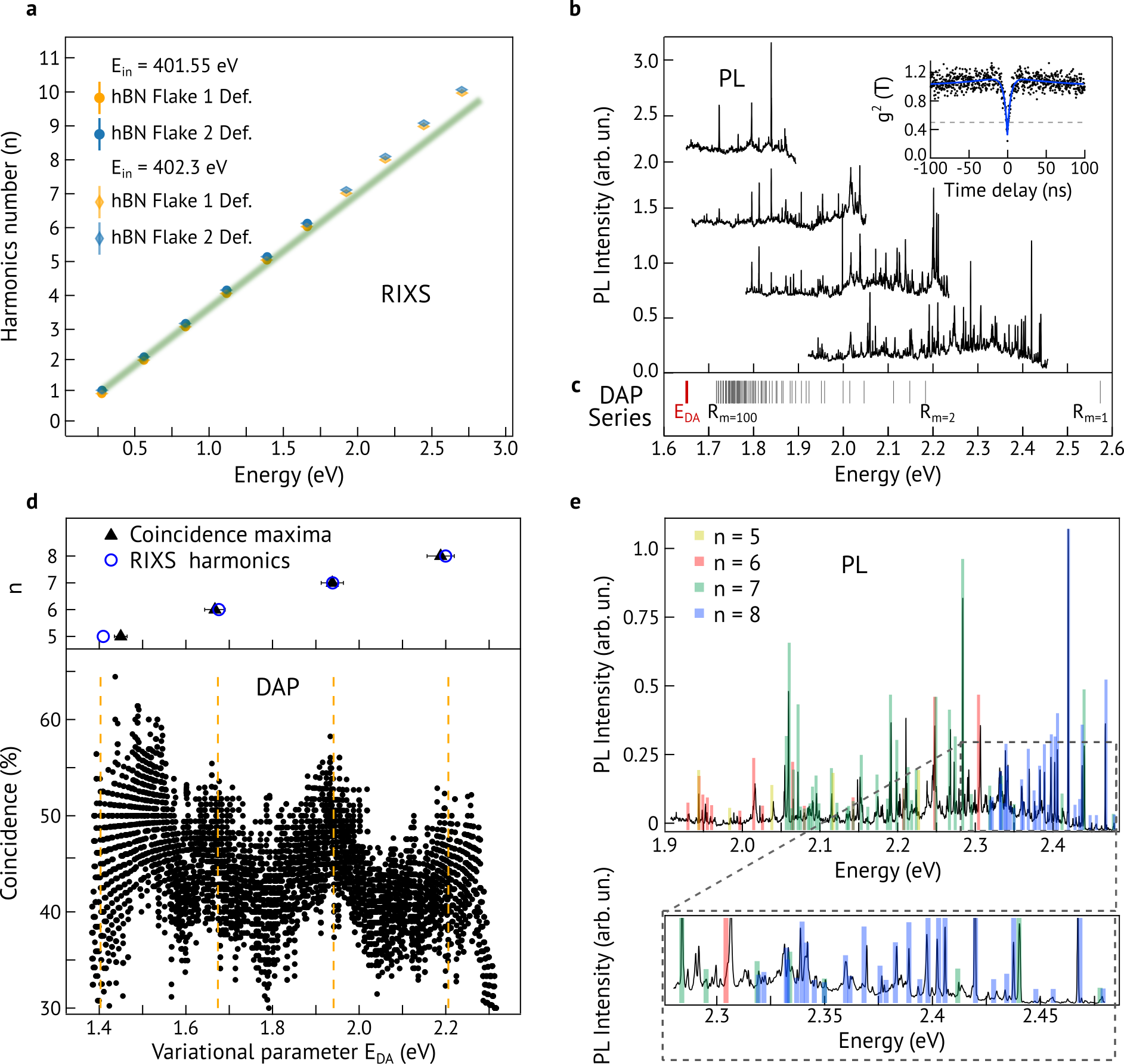}
\caption{ {\bf Fundamental energies in RIXS and PL experiments}. {\bf a} Dots indicate the energy of the harmonic peaks for two defective flakes at two incident energies. The error bars of the RIXS peak energies are established based on a combination of fitting, resolution, and estimation of the elastic line position.
{\bf b} Emission spectra of PL experiments at 8 K recorded in the same position of a defective sample with different excitation energy 
($E_{exc}$) and detection range ($E_{det}$): $E_{exc}=2.70$ eV and $E_{det}<2.48$ eV, $E_{exc}=2.43$ eV and $E_{det}<2.25$ eV, $E_{exc}=2.21$ eV and $E_{det}<2.07$ eV, $E_{exc}=2.03$ eV and $E_{det}<1.91$ eV. For clarity, spectra are displayed with vertical offset. The inset is an example of second-order autocorrelation function ($g^{(2)}(\tau)$) for a single line of the spectrum. {\bf c} Example of a DAP series ($\{E_m\}$) for $E_{DA}=1.65 eV$ up to $R_{m=100}$.
{\bf d} Coincidences between the SPE peaks in {\bf b} and the DAP series calculated with Eq.~\ref{eq1} by varying $E_{DA}$. Fundamental transitions $E^{i}_{DA}$ maximize coincidence counts and minimize the fit error (reported in SI). Top panel shows the values of $E^{i}_{DA}$ resulting from the DAP fitting vs the harmonic index of RIXS. Error bars correspond to the fit error. Dashed orange vertical lines indicate the energy of the harmonics measured with RIXS for $n=5,6,7,8$. {\bf e} PL spectrum for $E_{exc}=2.70$ eV overlaid to the matching lines (colored bars) of the DAP sequences from  Eq.~\ref{eq3}. The width of the colored bars indicate the matching tolerance of 2 meV, while the height does not have physical meaning and it is chosen to resemble the intensity of the corresponding peaks.}
\label{fig4}
\end{figure*}

We extract the energy, intensity, and linewidth of the RIXS peaks at the $N -\pi^*$ resonance by fitting them with Gaussian functions (see SI for further details).
In Fig.\ref{fig3}\textbf{d} we compare the peaks intensity of defective Flakes 1 and 2. In both cases, the intensity does not follow any exponential or monotonic trend but it has a maximum at $n=2$ when excited at $E_{in}=401.55$ eV whereas, for $E_{in}=402.3$ eV the trend seems to initially decrease and then increases again for $n=7$ and $n=8$. Previous theoretical works indicate that an exponential or monotonic decay is expected when electron-phonon coupling drives phonon detection in RIXS on solid state materials \cite{ament_determining_2011, geondzhian_demonstration_2018,geondzhian_generalization_2020, feng_disparate_2020,dashwood_probing_2021}. The non-exponential and non-monotonic trend observed in hBN suggests that these harmonics have a non-trivial electronic nature whose full understanding requires further dedicated studies. Similar non-exponential and non-monotonic trends have been reported for RIXS at the $N -\pi^*$ resonance in N$_2$ molecules, and have been associated with the vibrational excitation of the ground state of N$_2$ \cite{kjellsson2021resonant,lindblad2020x}. 
Furthermore, the energies of the harmonics exhibit an almost linear trend for all defective samples (Fig. \ref{fig4}\textbf{a} and Table \ref{table}), indicating a common intrinsic nature. 
There are multiple possible origins for the coherent modes emerging in defective hBN samples, but a large fraction of excitations can be ruled out by arguments based on the RIXS cross section, comparison with other solid state systems, and previous literature. Single spin excitations are excluded as they cannot be observed at the K edges because of the lack of spin-orbit coupling of the 1$s$ core electrons \cite{ament2011resonant}. 
Multi spin excitations can also be discarded as the number of harmonics (up to 10) is inconsistent with the spin state of nitrogen. There are also reports of hybridized \textit{dd} excitations coupled to phonons, but no harmonics were observed. We also exclude optical orbitons because we are not aware of any observations of excitons displaying a series of harmonics extending as high as the ones emerging in hBN.  \cite{lee2014charge,martinelli2023collective}. Moreover, excitons in hBN appear at an energy comparable to the band gap, and are not compatible with an elementary energy at 285 meV. Excitations connected with the Kekule' structure (such as Kekule' bond order excitations) of hBN can also be excluded as they should be present in pristine hBN.
Therefore, the origin of the harmonics should be researched in the defective structures that give rise to quantum emission.

To deepen our understanding of defective hBN, we compare the RIXS data with the results from PL experiments. 
First, we note that a vast amount of literature shows that SPEs are commonly found at around 580 nm (2.138 eV) and 640 nm (1.937 eV) \cite{comtet_2019_wide,hayee_2020_revealing,tran_2016_robust,mendelson_2019_engineering,grosso_tunable_2017,xu_single_2018}, with a striking match to the 7$^\textit{th}$ and 8$^\textit{th}$ peak of the harmonics in the RIXS data. Moreover, SPEs have been observed at other energies reached by the harmonics, including the ones in the NIR (1.25 eV) \cite{camphausen_observation_2020} and in the UV (2.8 eV) \cite{fournier_position-controlled_2021}. Figure \ref{fig4}\textbf{b} shows the PL emission spectra of a plasma-treated defective sample at 8 K obtained by laser excitation at different energies (see Methods and SI). We use different laser excitation energies to mitigate the dependence of SPEs on the incident energy and to access as many peaks as possible \cite{schell_quantum_2018,grosso_2020_low}. The quantum nature of the emission is confirmed by second-order autocorrelation function measurements as displayed in the inset of Fig. \ref{fig4}\textbf{b}. The PL emission spectra from defective hBN (Fig.\ref{fig4}\textbf{b}) cannot be explained by only the harmonics $E_{n}$ even considering the energy shifts due to strain or electromagnetic variations (in the order of 65  meV \cite{grosso_tunable_2017,mendelson_2020_strain} and 15 meV,\cite{nikolay_2019_very}, respectively). 
Further recombination mechanisms due to the incoherent nature of the PL decay process have therefore to be considered to account for the large spectral span of the PL emission from SPEs in defective hBN. 
Consequently, to compare the results of the coherent excitations of RIXS with the SPEs in PL experiments, we use a phenomenological model similar to the DAP recombination process that has been used to understand PL in many defective semiconductor materials and hBN \cite{Williams_1968_donor,tan_donoracceptor_2022}. 
A DAP process occurs when an electron from a donor defect recombines with an acceptor defect that can be at a distance of several atomic sites, as illustrated in Fig.~\ref{fig1}\textbf{b}. A single DAP recombination generates a PL emission peak with energy $E_{DA} + \frac{e^2}{4\pi\epsilon R}$, 
where $R$ is the spatial separation between the donor and acceptor site, $\epsilon=\epsilon_0\epsilon_{hBN}$ is the in-plane, bulk dielectric constant \cite{laturia2018dielectric}, and $E_{DA}$ is the energy difference between the energetic levels of the donor and acceptor defects. In the presence of several defects, recombination can occur among donors and acceptors at different sites of the lattice, and the DAP process generates a sequence of emission peaks following the discrete series:
\begin{equation}
\{E_{DA} + \frac{e^2}{4\pi\epsilon R_m}, \quad m \in \mathbb{N}\} 
\label{eq1}
\end{equation}
where $m$ is the index over all possible distances of the lattice sites ($R_m$). To account for the layered structure of hBN, we include all possible distances among lattice sites across two adjacent hBN layers, with $R_m = \{|i\vec{a}+j\vec{b}+k\vec{c}+l\vec{d}|, i,j \in \mathbb{Z} \quad \textrm{and} \quad k,l =0,1\}$, where $\vec{a}, \vec{b}, \vec{c}$, and $\vec{d}$ are the lattice vectors as defined in Fig.\ref{fig1}\textbf{a}.
An example of a DAP series in hBN for $E_{DA}= 1.65$ eV is displayed in Fig.~\ref{fig4}\textbf{c}. To analyze the PL emission of defective hBN, we compare the peak energies of the PL data of Fig.~\ref{fig4}\textbf{b} with the energy series of Eq.~\ref{eq1} generated by varying $E_{DA}$. We look for fundamental DAP transitions ($E^{i}_{DA}$) by estimating the number of peaks in the DAP series that match with the peaks in our data (coincidence counts) and evaluating the total energy difference between the DAP peaks sequence and the experimental PL values. The fit error is calculated from the total energy difference using the least squares method (see SI for a description of the procedure). Figure \ref{fig4}\textbf{d} reports the coincidence counts as a function of the variational parameter $E_{DA}$. The DAP analysis reveals the presence of four fundamental transitions $E^{i}_{DA}$ characterized by maxima of coincidence counts. These four peaks indicate that the system can host several donor-acceptor-like transitions whose energies ($E^{i}_{DA}$) are in excellent agreement with previous PL experiments on hBN samples of different origin. \cite{tan_donoracceptor_2022} 
Remarkably, the values of $E^{i}_{DA}$ exhibit a periodic energy spacing with a linear trend (see top panel of Fig.~\ref{fig4}\textbf{d}), and a notable match with the 5$^\textit{th}$, 6$^\textit{th}$, 7$^\textit{th}$, and 8$^\textit{th}$ harmonics observed in RIXS. The linear trend, the comparable energy values, and the regular spacing (of $\sim$ 285 meV) of $E^{i}_{DA}$ infer a correlation between the RIXS harmonic states ($E_n$) and SPEs. 
The results of the DAP analysis of PL spectra from different samples, reported in the SI, returns a common trend. The reliability of our DAP analysis is confirmed by a number of tests run on randomly simulated PL spectra and with synthetic DAP sequences summarized in the SI. 
A confidence interval for the values of the fundamental resonances $E^{i}_{DA}$ is provided by the fit error calculated from the total energy detuning between the ideal DAP sequence and the PL data. In Table \ref{table} we compare the results of RIXS and PL experiments together with values extracted from the literature in which a statistically significant number of zero-phonon lines (ZPL) transitions for SPEs in hBN are considered.

Having proved a periodic energy scale in the PL data, we now formulate an SPE series combining the RIXS harmonics and the DAP model: 

\begin{equation}
E_{SPE}=\{E^{i}_{DA} + \frac{e^2}{4\pi\epsilon R_m}\}_{i,m}=\{nE_{0} + \frac{e^2}{4\pi\epsilon R_m}\}_{n,m}
\label{eq3}
\end{equation}

where the energies of the harmonics ($E_{n}=nE_{0}$) play the role of the energy of the donor-acceptor transitions $E_{DA}$ of Eq.~\ref{eq1}. In Fig.~\ref{fig4}\textbf{e} we show that the series calculated with Eq.~\ref{eq3} can account for over 75$\%$ of the peaks observed in a PL spectrum of defective hBN. In the figure, we overlaid to a PL spectrum the lines of Eq.~\ref{eq3} that find a match with the experimental data. The width of the lines indicate the tolerance range of 2 meV used to find matching values between the calculated series and the PL spectrum. The full description of this procedure is reported in the SI. The good agreement between the PL spectra and the calculated peak sequences generated using the energy values of the harmonics emerging from the $N$-$\pi^*$ orbitals suggests that most of the SPEs observed in PL have a common origin associated with the elementary excitation $E_{0}$.

Although unable to offer a comprehensive understanding of the microscopic origin of SPEs in hBN, the DAP framework allows us to associate the elementary energy observed in RIXS with the single-photon emission process. Current theoretical approaches that aim to microscopically describe SPEs in hBN are mostly guided by the energy of ZPLs and phonon sidebands measured with PL spectroscopy.\cite{weston2018native,abdi_color_2018, tawfik_first-principles_2017, jungwirth_optical_2017,mendelson_identifying_2021} This often involves considering multiple SPEs independently, limiting the generality of the description, as we have proven that many SPEs are connected by an energy scale much lower than expected. Our evidence allows us to restrict the theoretical search for specific defects and highlights the importance of $p_z$ orbitals of nitrogen. Finally, we note that the similarities between the RIXS spectrum at the $N - \pi^*$ resonance of hBN and N$_2$ molecules could indicate a participation of N-N-like structures in the generation of SPEs.\cite{kjellsson2021resonant,lindblad2020x} 
In this regard, theoretical calculations reported that nitrogen interstitial defects incorporate in hBN in a split-interstitial configuration and result in a structure similar to a N$_2$ molecule introducing unoccupied gap states behaving as acceptors.\cite{weston2018native} In the future, the use of ab-initio theoretical methods (i.e. density functional theory, CCSD, quantum chemistry) that consider the presence of the harmonics states detected by RIXS and the involvement of N-$\pi^*$ orbitals will be fundamental for a deeper comprehension of quantum emission in hBN.  
Finally, we note that our findings do not exclude the presence in hBN of SPEs and fluorescent spin defects not related to harmonics at the N-$\pi^*$ resonance such as the $V^{-}_{B}$ defect.\cite{gottscholl_2021_room,gottscholl_2020_initialization}

In conclusion, by using RIXS in defective hBN we uncover the elementary energy ($E_0$) of a harmonic series which spans from the mid-IR to the UV, and correlates to SPEs observed in PL spectra in the visible range. 
Due to the resonance selectivity of RIXS we can pinpoint that N-$\pi^*$ bonds, rather than $\sigma^*$ bonds, are active in the electronic transitions responsible of single-photon emission. 
Overall, our evidence describes a large number of phenomena reported in the literature which are still under debate and sets solid basis for a complete description of single-photon emitters in hBN. Finally, our work establishes RIXS as an important tool in the discovery and comprehension of low-dimensional photonic quantum materials.

\newpage

\newpage

\clearpage
\begin{table}[!]
\begin{tabular}{ |p{1cm}|p{2.5cm}|p{3cm}|p{5.5cm}|  }
\hline
\multicolumn{4}{|c|}{Energy (in eV) of fundamental transitions as detected by different methods} \\
\hline
\rowcolor{yellow!50}[0pt][0pt] \emph{n} & RIXS Def. & PL & Other works \\
\hline
\rowcolor{black!10}[0pt][0pt] 1  & 0.285$\pm$ 0.002 & &  \\ 
\rowcolor{black!20}[0pt][0pt] 2  & 0.572$\pm$ 0.002 & & \\ 
\rowcolor{black!10}[0pt][0pt] 3  & 0.851$\pm$ 0.002 & & \\
\rowcolor{black!20}[0pt][0pt] 4  & 1.129$\pm$ 0.002 & & 1.240\cite{camphausen_observation_2020} \\ 
\rowcolor{black!10}[0pt][0pt] 5  & 1.403$\pm$ 0.002 & 1.444 $\pm$ 0.010 & 1.430\cite{tan_donoracceptor_2022} \\
\rowcolor{black!20}[0pt][0pt] 6  & 1.674$\pm$ 0.002 & 1.665 $\pm$ 0.024 & 1.708\cite{tan_donoracceptor_2022}, 1.759\cite{hayee_revealing_2020} \\ 
\rowcolor{black!10}[0pt][0pt] 7  & 1.941$\pm$ 0.002 & 1.940 $\pm$ 0.026 & 1.950\cite{tan_donoracceptor_2022}, 1.938\cite{comtet_2019_wide}, 2.033\cite{hayee_revealing_2020} \\
\rowcolor{black!20}[0pt][0pt] 8  & 2.206$\pm$ 0.002 & 2.194 $\pm$ 0.032 & 2.180\cite{tan_donoracceptor_2022}, 2.120\cite{comtet_2019_wide}, 2.131\cite{hayee_revealing_2020}, 2.138\cite{mendelson_2019_engineering} \\
\rowcolor{black!10}[0pt][0pt] 9  & 2.467$\pm$ 0.002 &  & 2.411\cite{tan_donoracceptor_2022} \\
\rowcolor{black!20}[0pt][0pt] 10 & 2.722$\pm$ 0.002 &  & 2.847\cite{fournier_position-controlled_2021} \\
\hline
\end{tabular}
\caption{{\bf Comparison between energy values in RIXS and PL.} The table compares the energy of the harmonics measured with RIXS, of the fundamental transitions extracted from the DAP model applied to PL spectra, and of the values of ZPL for SPEs in hBN reported in previous works in which large ensembles of emitters have been investigated. The error in the evaluation of the RIXS energies is based on an analysis of the fitting uncertainty, energy resolution, and identification of the elastic line. The energies extracted from the DAP analysis of the PL data are the average between the values obtained from the curves of the coincidence and the fit error as shown in Fig.\ref{fig4}\textbf{c}. The PL error is the error of the fit. }
\label{table}
\end{table}


\newpage

\section{Methods}
\paragraph{Growth and sample preparation:} 
High-quality single crystal hBN is grown under high pressure and high temperature.\cite{watanabe_2004_direct} Flakes of different thickness are obtained by mechanical exfoliation on Si/SiO$_2$ substrates. Defects are induced in pristine samples by irradiation with argon plasma inside a Reactive Ion Etcher (RIE) at 25 W for 5 minutes, followed by annealing at 950 $^{\circ}$C for 2 hours in a tube furnace purged with nitrogen.

\paragraph{XAS and RIXS measurements} 
The X-ray absorption (XAS) at the N-K edge was performed at the 2ID-SIX beamline at NSLS-II, Brookhaven National Laboratory (USA). The spectra were collected in total electron yield (TEY) and total fluorescence yield (TFY) at 300 K, using the vertically ($\sigma$) polarized light. The grazing incident angle is fixed at 30 degree for the XAS measurements. The X-ray beam size is 2x20 $\mu$m (VxH) and it is comparable to the flake size or smaller. Considering the size of the x-ray beam and the size of the flake we can exclude any contribution from the edges of the samples.

Resonant Inelastic X-ray Scattering (RIXS) measurements were performed at the 2ID-SIX beamline at NSLS-II, Brookhaven National Laboratory (USA). All samples are aligned with the surface normal (001) parallel to the scattering plane. The spectrometer arm is positioned at a fixed scattering angle 90 degrees. The incident light is $\sigma$-polarized. The energy resolution is about 20 meV (FWHM) at the N-K edge. All measurements are performed at 300 K. The fitting procedure is described in the supplementary information. The X-ray beam size is 2x20 $\mu$m (VxH) and it is comparable to the flake size or smaller. Considering the size of the x-ray beam and the size of the flake we can exclude any contribution from the edges of the samples.

\paragraph{PL measurements}
Photoluminescence (PL) and spectroscopy measurements are performed in a home-built microscope setup coupled to a closed-cycle cryostat. An objective with numerical aperture (NA) of 0.9 allows us to efficiently collect emission from areas with diameter of ~1.5 $\mu$m. Experiments are performed in a reflection geometry by exciting the sample with either a continuous-wave (CW) green laser (532 nm) or a supercontinuum pulsed laser with a tunable filter with a bandwidth of 2 nm. The laser reflection is removed from the PL signal by Long Pass filters.  Spectra are measured by a spectrometer with a high-resolution grating (1200 G/mm) and an EM-CCD camera. Second order correlation functions $g^{(2)}(\tau)$ are measured in a Hanbury-Brown and Twiss with free-space-coupled avalanche photodiodes (APDs) and by filtering individual spectral lines by tunable long and short pass filters.

\section{Acknowledgements} 
Work at Brookhaven National Laboratory was supported by the DOE Office of Science under Contract No.~DE-SC0012704. This work was also supported by the Laboratory Directed Research and Development project of  Brookhaven National Laboratory No.~21-037 and by the U.S.~Department of Energy (DOE) Office of Science, Early Career Research Program. This research used Beamline 2-ID of the National Synchrotron Light Source II, a U.S.~Department of Energy (DOE) Office of Science User Facility operated for the DOE Office of Science by Brookhaven National Laboratory under Contract No.~DE-SC0012704. Work at CUNY is supported by the National Science Foundation (NSF) (Grant No. DMR-2044281), by the physics department of the Graduate Center of CUNY and the Advanced Science Research Center, and support from the Research Foundation through PSC–CUNY Award No. 64510-00 52. K.W. and T.T. acknowledge support from the Elemental Strategy Initiative conducted by the MEXT, Japan (Grant No. JPMXP0112101001), and JSPS KAKENHI (Grant Nos. 19H05790, 20H00354, and 21H05233).



\bibliography{references2.bib}

\begin{thebibliography}{10}
\expandafter\ifx\csname url\endcsname\relax
  \def\url#1{\texttt{#1}}\fi
\expandafter\ifx\csname urlprefix\endcsname\relax\def\urlprefix{URL }\fi
\providecommand{\bibinfo}[2]{#2}
\providecommand{\eprint}[2][]{\url{#2}}

\bibitem{caldwell_photonics_2019}
\bibinfo{author}{Caldwell, J.~D.} \emph{et~al.}
\newblock \bibinfo{title}{Photonics with hexagonal boron nitride}.
\newblock \emph{\bibinfo{journal}{Nature Reviews Materials}} \textbf{\bibinfo{volume}{4}}, \bibinfo{pages}{552--567} (\bibinfo{year}{2019}).
\newblock \urlprefix\url{https://www.nature.com/articles/s41578-019-0124-1}.
\newblock \bibinfo{note}{Number: 8 Publisher: Nature Publishing Group}.

\bibitem{tran_quantum_2016}
\bibinfo{author}{Tran, T.~T.}, \bibinfo{author}{Bray, K.}, \bibinfo{author}{Ford, M.~J.}, \bibinfo{author}{Toth, M.} \& \bibinfo{author}{Aharonovich, I.}
\newblock \bibinfo{title}{Quantum emission from hexagonal boron nitride monolayers}.
\newblock \emph{\bibinfo{journal}{Nature Nanotechnology}} \textbf{\bibinfo{volume}{11}}, \bibinfo{pages}{37--41} (\bibinfo{year}{2016}).
\newblock \urlprefix\url{https://www.nature.com/articles/nnano.2015.242}.
\newblock \bibinfo{note}{Number: 1 Publisher: Nature Publishing Group}.

\bibitem{bourrellier_bright_2016}
\bibinfo{author}{Bourrellier, R.} \emph{et~al.}
\newblock \bibinfo{title}{Bright {UV} {Single} {Photon} {Emission} at {Point} {Defects} in h-{BN}}.
\newblock \emph{\bibinfo{journal}{Nano Letters}} \textbf{\bibinfo{volume}{16}}, \bibinfo{pages}{4317--4321} (\bibinfo{year}{2016}).
\newblock \urlprefix\url{https://doi.org/10.1021/acs.nanolett.6b01368}.
\newblock \bibinfo{note}{Publisher: American Chemical Society}.

\bibitem{grosso_tunable_2017}
\bibinfo{author}{Grosso, G.} \emph{et~al.}
\newblock \bibinfo{title}{Tunable and high-purity room temperature single-photon emission from atomic defects in hexagonal boron nitride}.
\newblock \emph{\bibinfo{journal}{Nature Communications}} \textbf{\bibinfo{volume}{8}}, \bibinfo{pages}{705} (\bibinfo{year}{2017}).
\newblock \urlprefix\url{https://www.nature.com/articles/s41467-017-00810-2}.
\newblock \bibinfo{note}{Number: 1 Publisher: Nature Publishing Group}.

\bibitem{fournier_position-controlled_2021}
\bibinfo{author}{Fournier, C.} \emph{et~al.}
\newblock \bibinfo{title}{Position-controlled quantum emitters with reproducible emission wavelength in hexagonal boron nitride}.
\newblock \emph{\bibinfo{journal}{Nature Communications}} \textbf{\bibinfo{volume}{12}}, \bibinfo{pages}{3779} (\bibinfo{year}{2021}).
\newblock \urlprefix\url{https://www.nature.com/articles/s41467-021-24019-6}.
\newblock \bibinfo{note}{Bandiera\_abtest: a Cc\_license\_type: cc\_by Cg\_type: Nature Research Journals Number: 1 Primary\_atype: Research Publisher: Nature Publishing Group Subject\_term: Single photons and quantum effects;Two-dimensional materials Subject\_term\_id: single-photons-and-quantum-effects;two-dimensional-materials}.

\bibitem{dietrich2018observation}
\bibinfo{author}{Dietrich, A.} \emph{et~al.}
\newblock \bibinfo{title}{Observation of fourier transform limited lines in hexagonal boron nitride}.
\newblock \emph{\bibinfo{journal}{Physical Review B}} \textbf{\bibinfo{volume}{98}}, \bibinfo{pages}{081414} (\bibinfo{year}{2018}).

\bibitem{chejanovsky2021single}
\bibinfo{author}{Chejanovsky, N.} \emph{et~al.}
\newblock \bibinfo{title}{Single-spin resonance in a van der waals embedded paramagnetic defect}.
\newblock \emph{\bibinfo{journal}{Nature materials}} \textbf{\bibinfo{volume}{20}}, \bibinfo{pages}{1079--1084} (\bibinfo{year}{2021}).

\bibitem{jungwirth_optical_2017}
\bibinfo{author}{Jungwirth, N.~R.} \& \bibinfo{author}{Fuchs, G.~D.}
\newblock \bibinfo{title}{Optical {Absorption} and {Emission} {Mechanisms} of {Single} {Defects} in {Hexagonal} {Boron} {Nitride}}.
\newblock \emph{\bibinfo{journal}{Physical Review Letters}} \textbf{\bibinfo{volume}{119}}, \bibinfo{pages}{057401} (\bibinfo{year}{2017}).
\newblock \urlprefix\url{https://link.aps.org/doi/10.1103/PhysRevLett.119.057401}.
\newblock \bibinfo{note}{Publisher: American Physical Society}.

\bibitem{abdi_color_2018}
\bibinfo{author}{Abdi, M.}, \bibinfo{author}{Chou, J.-P.}, \bibinfo{author}{Gali, A.} \& \bibinfo{author}{Plenio, M.~B.}
\newblock \bibinfo{title}{Color {Centers} in {Hexagonal} {Boron} {Nitride} {Monolayers}: {A} {Group} {Theory} and {Ab} {Initio} {Analysis}}.
\newblock \emph{\bibinfo{journal}{ACS Photonics}} \textbf{\bibinfo{volume}{5}}, \bibinfo{pages}{1967--1976} (\bibinfo{year}{2018}).
\newblock \urlprefix\url{https://doi.org/10.1021/acsphotonics.7b01442}.
\newblock \bibinfo{note}{Publisher: American Chemical Society}.

\bibitem{tawfik_first-principles_2017}
\bibinfo{author}{Tawfik, S.~A.} \emph{et~al.}
\newblock \bibinfo{title}{First-principles investigation of quantum emission from {hBN} defects}.
\newblock \emph{\bibinfo{journal}{Nanoscale}} \textbf{\bibinfo{volume}{9}}, \bibinfo{pages}{13575--13582} (\bibinfo{year}{2017}).
\newblock \urlprefix\url{https://pubs.rsc.org/en/content/articlelanding/2017/nr/c7nr04270a}.
\newblock \bibinfo{note}{Publisher: The Royal Society of Chemistry}.

\bibitem{mendelson_identifying_2021}
\bibinfo{author}{Mendelson, N.} \emph{et~al.}
\newblock \bibinfo{title}{Identifying carbon as the source of visible single-photon emission from hexagonal boron nitride}.
\newblock \emph{\bibinfo{journal}{Nature Materials}} \textbf{\bibinfo{volume}{20}}, \bibinfo{pages}{321--328} (\bibinfo{year}{2021}).
\newblock \urlprefix\url{https://www.nature.com/articles/s41563-020-00850-y}.
\newblock \bibinfo{note}{Number: 3 Publisher: Nature Publishing Group}.

\bibitem{gottscholl_2021_room}
\bibinfo{author}{Gottscholl, A.} \emph{et~al.}
\newblock \bibinfo{title}{Room temperature coherent control of spin defects in hexagonal boron nitride}.
\newblock \emph{\bibinfo{journal}{Science Advances}} \textbf{\bibinfo{volume}{7}}, \bibinfo{pages}{eabf3630} (\bibinfo{year}{2021}).

\bibitem{gottscholl_2020_initialization}
\bibinfo{author}{Gottscholl, A.} \emph{et~al.}
\newblock \bibinfo{title}{Initialization and read-out of intrinsic spin defects in a van der waals crystal at room temperature}.
\newblock \emph{\bibinfo{journal}{Nature materials}} \textbf{\bibinfo{volume}{19}}, \bibinfo{pages}{540--545} (\bibinfo{year}{2020}).

\bibitem{tran_2016_robust}
\bibinfo{author}{Tran, T.~T.} \emph{et~al.}
\newblock \bibinfo{title}{Robust multicolor single photon emission from point defects in hexagonal boron nitride}.
\newblock \emph{\bibinfo{journal}{ACS nano}} \textbf{\bibinfo{volume}{10}}, \bibinfo{pages}{7331--7338} (\bibinfo{year}{2016}).

\bibitem{ament_resonant_2011}
\bibinfo{author}{Ament, L. J.~P.}, \bibinfo{author}{van Veenendaal, M.}, \bibinfo{author}{Devereaux, T.~P.}, \bibinfo{author}{Hill, J.~P.} \& \bibinfo{author}{van~den Brink, J.}
\newblock \bibinfo{title}{Resonant inelastic x-ray scattering studies of elementary excitations}.
\newblock \emph{\bibinfo{journal}{Reviews of Modern Physics}} \textbf{\bibinfo{volume}{83}}, \bibinfo{pages}{705--767} (\bibinfo{year}{2011}).
\newblock \urlprefix\url{http://link.aps.org/doi/10.1103/RevModPhys.83.705}.

\bibitem{ament_determining_2011}
\bibinfo{author}{Ament, L. J.~P.}, \bibinfo{author}{van Veenendaal, M.} \& \bibinfo{author}{van~den Brink, J.}
\newblock \bibinfo{title}{Determining the electron-phonon coupling strength from {Resonant} {Inelastic} {X}-ray {Scattering} at transition metal {L}-edges}.
\newblock \emph{\bibinfo{journal}{EPL (Europhysics Letters)}} \textbf{\bibinfo{volume}{95}}, \bibinfo{pages}{27008} (\bibinfo{year}{2011}).
\newblock \urlprefix\url{https://iopscience.iop.org/article/10.1209/0295-5075/95/27008}.

\bibitem{gelmukhanov_dynamics_2021}
\bibinfo{author}{Gel’mukhanov, F.}, \bibinfo{author}{Odelius, M.}, \bibinfo{author}{Polyutov, S.~P.}, \bibinfo{author}{Föhlisch, A.} \& \bibinfo{author}{Kimberg, V.}
\newblock \bibinfo{title}{Dynamics of resonant x-ray and {Auger} scattering}.
\newblock \emph{\bibinfo{journal}{Reviews of Modern Physics}} \textbf{\bibinfo{volume}{93}}, \bibinfo{pages}{035001} (\bibinfo{year}{2021}).
\newblock \urlprefix\url{https://link.aps.org/doi/10.1103/RevModPhys.93.035001}.

\bibitem{dvorak_towards_2016}
\bibinfo{author}{Dvorak, J.}, \bibinfo{author}{Jarrige, I.}, \bibinfo{author}{Bisogni, V.}, \bibinfo{author}{Coburn, S.} \& \bibinfo{author}{Leonhardt, W.}
\newblock \bibinfo{title}{Towards 10 {meV} resolution: {The} design of an ultrahigh resolution soft {X}-ray {RIXS} spectrometer}.
\newblock \emph{\bibinfo{journal}{Review of Scientific Instruments}} \textbf{\bibinfo{volume}{87}}, \bibinfo{pages}{115109} (\bibinfo{year}{2016}).
\newblock \urlprefix\url{http://aip.scitation.org/doi/10.1063/1.4964847}.

\bibitem{pelliciari_tuning_2021}
\bibinfo{author}{Pelliciari, J.} \emph{et~al.}
\newblock \bibinfo{title}{Tuning spin excitations in magnetic films by confinement}.
\newblock \emph{\bibinfo{journal}{Nature Materials}} \textbf{\bibinfo{volume}{20}}, \bibinfo{pages}{188--193} (\bibinfo{year}{2021}).
\newblock \urlprefix\url{https://www.nature.com/articles/s41563-020-00878-0}.
\newblock \bibinfo{note}{Number: 2 Publisher: Nature Publishing Group}.

\bibitem{pelliciari_evolution_2021}
\bibinfo{author}{Pelliciari, J.} \emph{et~al.}
\newblock \bibinfo{title}{Evolution of spin excitations from bulk to monolayer {FeSe}}.
\newblock \emph{\bibinfo{journal}{Nature Communications}} \textbf{\bibinfo{volume}{12}}, \bibinfo{pages}{3122} (\bibinfo{year}{2021}).
\newblock \urlprefix\url{https://www.nature.com/articles/s41467-021-23317-3}.
\newblock \bibinfo{note}{Number: 1 Publisher: Nature Publishing Group}.

\bibitem{dean_spin_2012}
\bibinfo{author}{Dean, M. P.~M.} \emph{et~al.}
\newblock \bibinfo{title}{Spin excitations in a single {La2CuO4} layer}.
\newblock \emph{\bibinfo{journal}{Nature Materials}} \textbf{\bibinfo{volume}{11}}, \bibinfo{pages}{850--854} (\bibinfo{year}{2012}).
\newblock \urlprefix\url{http://www.nature.com/nmat/journal/v11/n10/full/nmat3409.html}.

\bibitem{mcdougall_influence_2017}
\bibinfo{author}{McDougall, N.~L.}, \bibinfo{author}{Partridge, J.~G.}, \bibinfo{author}{Nicholls, R.~J.}, \bibinfo{author}{Russo, S.~P.} \& \bibinfo{author}{McCulloch, D.~G.}
\newblock \bibinfo{title}{Influence of point defects on the near edge structure of hexagonal boron nitride}.
\newblock \emph{\bibinfo{journal}{Physical Review B}} \textbf{\bibinfo{volume}{96}}, \bibinfo{pages}{144106} (\bibinfo{year}{2017}).
\newblock \urlprefix\url{https://link.aps.org/doi/10.1103/PhysRevB.96.144106}.
\newblock \bibinfo{note}{Publisher: American Physical Society}.

\bibitem{vinson_resonant_2017}
\bibinfo{author}{Vinson, J.}, \bibinfo{author}{Jach, T.}, \bibinfo{author}{Müller, M.}, \bibinfo{author}{Unterumsberger, R.} \& \bibinfo{author}{Beckhoff, B.}
\newblock \bibinfo{title}{Resonant x-ray emission of hexagonal boron nitride}.
\newblock \emph{\bibinfo{journal}{Physical Review B}} \textbf{\bibinfo{volume}{96}}, \bibinfo{pages}{205116} (\bibinfo{year}{2017}).
\newblock \urlprefix\url{https://link.aps.org/doi/10.1103/PhysRevB.96.205116}.

\bibitem{mcdougall_near_2014}
\bibinfo{author}{McDougall, N.~L.}, \bibinfo{author}{Nicholls, R.~J.}, \bibinfo{author}{Partridge, J.~G.} \& \bibinfo{author}{McCulloch, D.~G.}
\newblock \bibinfo{title}{The {Near} {Edge} {Structure} of {Hexagonal} {Boron} {Nitride}}.
\newblock \emph{\bibinfo{journal}{Microscopy and Microanalysis}} \textbf{\bibinfo{volume}{20}}, \bibinfo{pages}{1053--1059} (\bibinfo{year}{2014}).
\newblock \urlprefix\url{https://www.cambridge.org/core/product/identifier/S1431927614000737/type/journal_article}.

\bibitem{watanabe2004direct}
\bibinfo{author}{Watanabe, K.}, \bibinfo{author}{Taniguchi, T.} \& \bibinfo{author}{Kanda, H.}
\newblock \bibinfo{title}{Direct-bandgap properties and evidence for ultraviolet lasing of hexagonal boron nitride single crystal}.
\newblock \emph{\bibinfo{journal}{Nature materials}} \textbf{\bibinfo{volume}{3}}, \bibinfo{pages}{404--409} (\bibinfo{year}{2004}).

\bibitem{xu_single_2018}
\bibinfo{author}{Xu, Z.-Q.} \emph{et~al.}
\newblock \bibinfo{title}{Single photon emission from plasma treated {2D} hexagonal boron nitride}.
\newblock \emph{\bibinfo{journal}{Nanoscale}} \textbf{\bibinfo{volume}{10}}, \bibinfo{pages}{7957--7965} (\bibinfo{year}{2018}).
\newblock \urlprefix\url{https://pubs.rsc.org/en/content/articlelanding/2018/nr/c7nr08222c}.
\newblock \bibinfo{note}{Publisher: The Royal Society of Chemistry}.

\bibitem{serrano_vibrational_2007}
\bibinfo{author}{Serrano, J.} \emph{et~al.}
\newblock \bibinfo{title}{Vibrational {Properties} of {Hexagonal} {Boron} {Nitride}: {Inelastic} {X}-{Ray} {Scattering} and \textit{{Ab} {Initio}} {Calculations}}.
\newblock \emph{\bibinfo{journal}{Physical Review Letters}} \textbf{\bibinfo{volume}{98}}, \bibinfo{pages}{095503} (\bibinfo{year}{2007}).
\newblock \urlprefix\url{https://link.aps.org/doi/10.1103/PhysRevLett.98.095503}.

\bibitem{geondzhian_demonstration_2018}
\bibinfo{author}{Geondzhian, A.} \& \bibinfo{author}{Gilmore, K.}
\newblock \bibinfo{title}{Demonstration of resonant inelastic x-ray scattering as a probe of exciton-phonon coupling}.
\newblock \emph{\bibinfo{journal}{Physical Review B}} \textbf{\bibinfo{volume}{98}}, \bibinfo{pages}{214305} (\bibinfo{year}{2018}).
\newblock \urlprefix\url{https://link.aps.org/doi/10.1103/PhysRevB.98.214305}.

\bibitem{geondzhian_generalization_2020}
\bibinfo{author}{Geondzhian, A.} \& \bibinfo{author}{Gilmore, K.}
\newblock \bibinfo{title}{Generalization of the {Franck}-{Condon} model for phonon excitations by resonant inelastic x-ray scattering}.
\newblock \emph{\bibinfo{journal}{Physical Review B}} \textbf{\bibinfo{volume}{101}}, \bibinfo{pages}{214307} (\bibinfo{year}{2020}).
\newblock \urlprefix\url{https://link.aps.org/doi/10.1103/PhysRevB.101.214307}.

\bibitem{feng_disparate_2020}
\bibinfo{author}{Feng, X.} \emph{et~al.}
\newblock \bibinfo{title}{Disparate {Exciton}-{Phonon} {Couplings} for {Zone}-{Center} and {Boundary} {Phonons} in {Solid}-{State} {Graphite}}.
\newblock \emph{\bibinfo{journal}{Physical Review Letters}} \textbf{\bibinfo{volume}{125}}, \bibinfo{pages}{116401} (\bibinfo{year}{2020}).
\newblock \urlprefix\url{https://link.aps.org/doi/10.1103/PhysRevLett.125.116401}.
\newblock \bibinfo{note}{Publisher: American Physical Society}.

\bibitem{dashwood_probing_2021}
\bibinfo{author}{Dashwood, C.} \emph{et~al.}
\newblock \bibinfo{title}{Probing {Electron}-{Phonon} {Interactions} {Away} from the {Fermi} {Level} with {Resonant} {Inelastic} {X}-{Ray} {Scattering}}.
\newblock \emph{\bibinfo{journal}{Physical Review X}} \textbf{\bibinfo{volume}{11}}, \bibinfo{pages}{041052} (\bibinfo{year}{2021}).
\newblock \urlprefix\url{https://link.aps.org/doi/10.1103/PhysRevX.11.041052}.

\bibitem{kjellsson2021resonant}
\bibinfo{author}{Kjellsson, L.} \emph{et~al.}
\newblock \bibinfo{title}{Resonant inelastic x-ray scattering at the n 2 $\pi$* resonance: Lifetime-vibrational interference, radiative electron rearrangement, and wave-function imaging}.
\newblock \emph{\bibinfo{journal}{Physical Review A}} \textbf{\bibinfo{volume}{103}}, \bibinfo{pages}{022812} (\bibinfo{year}{2021}).

\bibitem{lindblad2020x}
\bibinfo{author}{Lindblad, R.} \emph{et~al.}
\newblock \bibinfo{title}{X-ray absorption spectrum of the n 2+ molecular ion}.
\newblock \emph{\bibinfo{journal}{Physical Review Letters}} \textbf{\bibinfo{volume}{124}}, \bibinfo{pages}{203001} (\bibinfo{year}{2020}).

\bibitem{ament2011resonant}
\bibinfo{author}{Ament, L.~J.}, \bibinfo{author}{Van~Veenendaal, M.}, \bibinfo{author}{Devereaux, T.~P.}, \bibinfo{author}{Hill, J.~P.} \& \bibinfo{author}{Van Den~Brink, J.}
\newblock \bibinfo{title}{Resonant inelastic x-ray scattering studies of elementary excitations}.
\newblock \emph{\bibinfo{journal}{Reviews of Modern Physics}} \textbf{\bibinfo{volume}{83}}, \bibinfo{pages}{705} (\bibinfo{year}{2011}).

\bibitem{lee2014charge}
\bibinfo{author}{Lee, J.} \emph{et~al.}
\newblock \bibinfo{title}{Charge-orbital-lattice coupling effects in the d d excitation profile of one-dimensional cuprates}.
\newblock \emph{\bibinfo{journal}{Physical Review B}} \textbf{\bibinfo{volume}{89}}, \bibinfo{pages}{041104} (\bibinfo{year}{2014}).

\bibitem{martinelli2023collective}
\bibinfo{author}{Martinelli, L.} \emph{et~al.}
\newblock \bibinfo{title}{Collective nature of orbital excitations in layered cuprates in the absence of apical oxygens}.
\newblock \emph{\bibinfo{journal}{arXiv preprint arXiv:2304.02115}}  (\bibinfo{year}{2023}).

\bibitem{comtet_2019_wide}
\bibinfo{author}{Comtet, J.} \emph{et~al.}
\newblock \bibinfo{title}{Wide-field spectral super-resolution mapping of optically active defects in hexagonal boron nitride}.
\newblock \emph{\bibinfo{journal}{Nano letters}} \textbf{\bibinfo{volume}{19}}, \bibinfo{pages}{2516--2523} (\bibinfo{year}{2019}).

\bibitem{hayee_2020_revealing}
\bibinfo{author}{Hayee, F.} \emph{et~al.}
\newblock \bibinfo{title}{Revealing multiple classes of stable quantum emitters in hexagonal boron nitride with correlated optical and electron microscopy}.
\newblock \emph{\bibinfo{journal}{Nature materials}} \textbf{\bibinfo{volume}{19}}, \bibinfo{pages}{534--539} (\bibinfo{year}{2020}).

\bibitem{mendelson_2019_engineering}
\bibinfo{author}{Mendelson, N.} \emph{et~al.}
\newblock \bibinfo{title}{Engineering and tuning of quantum emitters in few-layer hexagonal boron nitride}.
\newblock \emph{\bibinfo{journal}{ACS nano}} \textbf{\bibinfo{volume}{13}}, \bibinfo{pages}{3132--3140} (\bibinfo{year}{2019}).

\bibitem{camphausen_observation_2020}
\bibinfo{author}{Camphausen, R.} \emph{et~al.}
\newblock \bibinfo{title}{Observation of near-infrared sub-{Poissonian} photon emission in hexagonal boron nitride at room temperature}.
\newblock \emph{\bibinfo{journal}{APL Photonics}} \textbf{\bibinfo{volume}{5}}, \bibinfo{pages}{076103} (\bibinfo{year}{2020}).
\newblock \urlprefix\url{https://aip.scitation.org/doi/10.1063/5.0008242}.
\newblock \bibinfo{note}{Publisher: American Institute of Physics}.

\bibitem{schell_quantum_2018}
\bibinfo{author}{Schell, A.~W.}, \bibinfo{author}{Svedendahl, M.} \& \bibinfo{author}{Quidant, R.}
\newblock \bibinfo{title}{Quantum {Emitters} in {Hexagonal} {Boron} {Nitride} {Have} {Spectrally} {Tunable} {Quantum} {Efficiency}}.
\newblock \emph{\bibinfo{journal}{Advanced Materials}} \textbf{\bibinfo{volume}{30}}, \bibinfo{pages}{1704237} (\bibinfo{year}{2018}).
\newblock \urlprefix\url{https://onlinelibrary.wiley.com/doi/10.1002/adma.201704237}.

\bibitem{grosso_2020_low}
\bibinfo{author}{Grosso, G.} \emph{et~al.}
\newblock \bibinfo{title}{Low-temperature electron--phonon interaction of quantum emitters in hexagonal boron nitride}.
\newblock \emph{\bibinfo{journal}{ACS Photonics}} \textbf{\bibinfo{volume}{7}}, \bibinfo{pages}{1410--1417} (\bibinfo{year}{2020}).

\bibitem{mendelson_2020_strain}
\bibinfo{author}{Mendelson, N.}, \bibinfo{author}{Doherty, M.}, \bibinfo{author}{Toth, M.}, \bibinfo{author}{Aharonovich, I.} \& \bibinfo{author}{Tran, T.~T.}
\newblock \bibinfo{title}{Strain-induced modification of the optical characteristics of quantum emitters in hexagonal boron nitride}.
\newblock \emph{\bibinfo{journal}{Advanced Materials}} \textbf{\bibinfo{volume}{32}}, \bibinfo{pages}{1908316} (\bibinfo{year}{2020}).

\bibitem{nikolay_2019_very}
\bibinfo{author}{Nikolay, N.} \emph{et~al.}
\newblock \bibinfo{title}{Very large and reversible stark-shift tuning of single emitters in layered hexagonal boron nitride}.
\newblock \emph{\bibinfo{journal}{Physical Review Applied}} \textbf{\bibinfo{volume}{11}}, \bibinfo{pages}{041001} (\bibinfo{year}{2019}).

\bibitem{Williams_1968_donor}
\bibinfo{author}{Williams, F.}
\newblock \bibinfo{title}{Donor—acceptor pairs in semiconductors}.
\newblock \emph{\bibinfo{journal}{physica status solidi (b)}} \textbf{\bibinfo{volume}{25}}, \bibinfo{pages}{493--512} (\bibinfo{year}{1968}).
\newblock \urlprefix\url{https://onlinelibrary.wiley.com/doi/abs/10.1002/pssb.19680250202}.
\newblock \eprint{https://onlinelibrary.wiley.com/doi/pdf/10.1002/pssb.19680250202}.

\bibitem{tan_donoracceptor_2022}
\bibinfo{author}{Tan, Q.} \emph{et~al.}
\newblock \bibinfo{title}{Donor–{Acceptor} {Pair} {Quantum} {Emitters} in {Hexagonal} {Boron} {Nitride}}.
\newblock \emph{\bibinfo{journal}{Nano Letters}} \textbf{\bibinfo{volume}{22}}, \bibinfo{pages}{1331--1337} (\bibinfo{year}{2022}).
\newblock \urlprefix\url{https://pubs.acs.org/doi/10.1021/acs.nanolett.1c04647}.

\bibitem{laturia2018dielectric}
\bibinfo{author}{Laturia, A.}, \bibinfo{author}{Van~de Put, M.~L.} \& \bibinfo{author}{Vandenberghe, W.~G.}
\newblock \bibinfo{title}{Dielectric properties of hexagonal boron nitride and transition metal dichalcogenides: from monolayer to bulk}.
\newblock \emph{\bibinfo{journal}{npj 2D Materials and Applications}} \textbf{\bibinfo{volume}{2}}, \bibinfo{pages}{1--7} (\bibinfo{year}{2018}).

\bibitem{weston2018native}
\bibinfo{author}{Weston, L.}, \bibinfo{author}{Wickramaratne, D.}, \bibinfo{author}{Mackoit, M.}, \bibinfo{author}{Alkauskas, A.} \& \bibinfo{author}{Van~de Walle, C.}
\newblock \bibinfo{title}{Native point defects and impurities in hexagonal boron nitride}.
\newblock \emph{\bibinfo{journal}{Physical Review B}} \textbf{\bibinfo{volume}{97}}, \bibinfo{pages}{214104} (\bibinfo{year}{2018}).

\bibitem{hayee_revealing_2020}
\bibinfo{author}{Hayee, F.} \emph{et~al.}
\newblock \bibinfo{title}{Revealing multiple classes of stable quantum emitters in hexagonal boron nitride with correlated optical and electron microscopy}.
\newblock \emph{\bibinfo{journal}{Nature Materials}} \textbf{\bibinfo{volume}{19}}, \bibinfo{pages}{534--539} (\bibinfo{year}{2020}).
\newblock \urlprefix\url{https://www.nature.com/articles/s41563-020-0616-9}.
\newblock \bibinfo{note}{Number: 5 Publisher: Nature Publishing Group}.

\bibitem{watanabe_2004_direct}
\bibinfo{author}{Watanabe, K.}, \bibinfo{author}{Taniguchi, T.} \& \bibinfo{author}{Kanda, H.}
\newblock \bibinfo{title}{Direct-bandgap properties and evidence for ultraviolet lasing of hexagonal boron nitride single crystal}.
\newblock \emph{\bibinfo{journal}{Nature materials}} \textbf{\bibinfo{volume}{3}}, \bibinfo{pages}{404--409} (\bibinfo{year}{2004}).

\end{thebibliography}
\end{document}